# A Gradient Tree Boosting based Approach to Rumor Detecting on Sina Weibo

**Xiao Xiong**, School of Computer Science and Engineering, University of Electronic Science and Technology of China
**Bo Yang**[*], School of Computer Science and Engineering, University of Electronic Science and Technology of China. [*]Corresponding author, yangbo@uestc.edu.cn
**Zhongfeng Kang**, School of Computer Science and Engineering, University of Electronic Science and Technology of China

Rumor detecting on microblogging platforms such as Sina Weibo is a crucial issue. Most existing rumor detecting algorithms require a lot of propagation data for model training, thus they do not have good detecting accuracy at the early stage after a rumor message is posted. In this paper, we propose to use gradient tree boosting (GTB) approach to rumor detecting, based on which a rumor detecting algorithm is developed. At the same time, the GTB-based approach makes it easy to conduct feature selection, and a feature selection algorithm is developed. Experiments on a widely used dataset of Sina Weibo show that the proposed detecting algorithm outperforms the state-of-the-art detecting algorithms; moreover, it has the highest detecting accuracy at the early stage. This work seems to be the first to use GTB-based approach in rumor detecting, and the results suggest that it may be a promising one.

## 1 INTRODUCTION

Microblogging platforms such as Sina Weibo[1] [1], similar to Twitter [2], allow users to instantly broadcast short messages with optional images and other kinds of multimedia data to all their followers. As one of the most influential social media sites in China, Weibo has more than 500 million users, and more than 100 million postings are posted on Weibo every day [18]. Weibo has a high speed of information propagation which makes it easy and fast for people to communicate with each other; on the other hand, Weibo also becomes an ideal platform for the spread of false information and fake news, i.e., rumors.

If a rumor is posted on Weibo, it can spread widely through all the followers. Incidents of online rumor spreading can lead to serious consequences such as massive public panic, social unrest, etc. For example, In July 2017 a rumor on Weibo claimed that there would be an earthquake within two months in Lianyungang[2], and many panic people who did not know the truth forwarded this message and soon it was diffused throughout the Internet. In the end, experts from the earthquake administration in China officially dispelled this rumor.

It can be noted that without proper background knowledge or apparent official evidence, it is rather difficult for people to distinguish between rumor messages and non-rumor messages. Therefore, it is desirable to automatically identify rumor messages on Weibo to prevent them from spreading, preferably as early as the messages are posted.

---

[1] In this paper, Sina Weibo is abbreviated as Weibo hereafter.
[2] A city in China.

In the literature, the problem of automatic detecting of rumors on Weibo has attracted substantial research. Most of related research uses classification algorithms for rumor detecting [3]-[10], among which the recurrent neural networks (RNN) based rumor detecting algorithm was shown to have the highest detecting accuracy [10]. Nevertheless, there is a main drawback to the state-of-the-art rumor detecting algorithms[3], i.e., the algorithms require a lot of propagation data for model training, thus satisfactory detecting accuracy can only be achieved when sufficient propagation data are available, which implies that the rumor message has already relatively widely spread. Since rumor detecting aims to identify rumors as early as possible to prevent them from spreading, unsatisfactory early detecting accuracy becomes a critical issue.

In this paper, we propose to use gradient tree boosting (GTB) approach in rumor detecting, based on which an algorithm, gradient tree boosting based rumor detecting (GTB-RD) algorithm is developed. The GTB-RD algorithm is an ensemble algorithm [11][12] that combines multiple base detectors to achieve high accuracy. The use of GTB-based approach has two desirable features. On the one hand, experiments on a widely used Weibo dataset suggest that the proposed GTB-RD algorithm seems to have the highest detecting accuracy, both at the early stage and in the long term, compared with the state-of-the-art detecting algorithms. For example, when detecting deadline is 0, i.e., as soon as a message is posted, the detecting accuracy is already 89.1%. On the other hand, GTB-based approach enables feature selection to be conducted in a convenient way, and a feature selection algorithm is developed in this paper. The use of proper features also contributes to the increase of detecting accuracy achieved by the GTB-RD algorithm. To the best of our knowledge, this work seems to be the first to try GTB-based approach on rumor detecting.

The remainder of this paper is organized as follows. In Section 2 related work is reviewed. In Section 3 the GTB-RD algorithm is proposed; and in Section 4 feature construction and selection for Weibo rumor detecting are conducted. Section 5 presents experimental studies and results. Section 6 concludes this paper.

## 2 RELATED WORK

Rumor detecting on social media began with the pioneering study of information credibility on Twitter [6]. Existing rumor detecting algorithms can be generally classified into three types: detecting by statistical features [4][6][7], detecting by feature variation over time [3][5][8][10], and detecting by messages' propagation structure [9]. Here we give a brief review on related work.

For rumor detecting by statistical features, most of existing work attempted to classify messages using information on message content, users and comments [4][6][7], and different sets of hand-crafted features were proposed and used.

For rumor detecting by feature variation over time, Zhao et al. [3] used clue terms such as "not true", "unconfirmed" or "debunk" to find questioning and denying posts; Ma et al. [5] and Kwon et al. [8] proposed time series algorithms for rumor detecting; Ma et al. [10] adopted deep learning algorithms, specifically, RNN-based algorithm, for rumor detecting. In [3][5][8][10], the variation of social context features during the message propagation over time was used.

For rumor detecting by messages' propagation structure, Wu et al. [9] proposed a graph-kernel-based hybrid SVM algorithm, which captured the propagation patterns as well as semantic features for rumor detecting. The propagation structure of a message was modeled by a propagation-tree-based structure.

Early rumor detecting is also an important issue, which has been studied in [3][5][10] by examining the variation of detecting accuracy with time. In this paper, we conduct similar study and compare the early detecting accuracy of the proposed GTB-RD algorithm with that of the algorithms in [3][5][10].

---

[3] including rumor detecting algorithms on Weibo and on other microblogging platforms such as Twitter.

# 3 A GTB-BASED APPROACH TO RUMOR DETECTING

## 3.1 Problem definition

Rumor detecting problem can be defined as follows (e.g., [3]-[10]). Consider a labeled dataset with $n$ posted messages:

$$\mathcal{D} = \{(x_1, y_1), (x_2, y_2), (x_3, y_3), \ldots, (x_n, y_n)\}, \tag{1}$$

where $x_i$ (i=1, 2..., n) represents the features of the $i$:th message and $y_i \in \{c_1, c_2\}$ (i=1, 2..., n) represents the label of message $x_i$. For each message $x_i$, there are $t$ features, $x_i^1, x_i^2, \ldots, x_i^t$. $c_1$ and $c_2$ represent a message is rumor or non-rumor, respectively. The goal of a rumor detecting algorithm is to give a predicted label for an unlabeled message, indicating whether the message is a rumor or not.

## 3.2 The Gradient Boosting Approach

Ensemble learning [11][12] combines multiple base classifiers to form a strong predicting model and generally obtains high predicting accuracy. One of the advantages of ensemble learning is that it does not require base classifiers to have high accuracy to obtain the overall high accuracy of the predicting model [19][20][21]. By ensemble learning, a complex problem can be decomposed into multiple sub-problems that are easier to solve and understand.

Boosting, e.g., gradient boosting [14][16], and bagging are two approaches commonly used in ensemble learning. In this paper, we propose to use gradient boosting approach, specifically, gradient tree boosting approach (details can be found in Section 3.3) to rumor detecting. Adopting gradient boosting approach, the output prediction $F_M(x_i)$ for message $x_i$ is the additive sum of $M$ base classifiers, referred to as base detectors in this paper, i.e.,

$$F_M(x_i) = \sum_{m=1}^{M} \gamma_m h_m(x_i), \tag{2}$$

where $h_m(x)$'s are base detectors in the context of boosting, $\gamma_m$ is the weight of detector $h_m(x)$. The above additive model is built in a forward stage-wise manner, i.e.,

$$F_m(x) = F_{m-1}(x) + \alpha \gamma_m h_m(x), m=1, 2,\ldots,M, \tag{3}$$

where $\alpha$ is the learning rate ($0<\alpha<=1$).

For stage $m$, a base detector $h_m(x)$ is chosen to be the one that minimizes the loss function $L$ given the current model $F_{m-1}(x)$ and its prediction, i.e.,

$$F_m(x) = F_{m-1}(x) + arg \min_h \sum_{i=1}^{n} L(y_i, F_{m-1}(x_i) + h(x_i)), \tag{4}$$

where $L$ is the mean square error, defined as

$$L(F_m(x), y) = \frac{1}{n}\sum_{i=1}^{n}(F_m(x_i) - y_i)^2 \tag{5}$$

The minimization problem can be numerically solved via negative gradient direction of the loss function, i.e.,

$$F_m(x) = F_{m-1}(x) - \gamma_m \sum_{i=1}^{n} \nabla_F L(y_i, F_{m-1}(x_i)) \tag{6}$$

In stage $m$, the predicting target for the base detector to fit is calculated by

$$z_{m,i} = -\left[\frac{\partial L(y_i, F_{m-1}(x_i))}{\partial F_{m-1}(x_i)}\right] \tag{7}$$

The weight of base detector is calculated by

$$\gamma_m = arg \min_\gamma \sum_{i=1}^{n} L\left(y_i, F_{m-1}(x_i) - \gamma \frac{\partial L(y_i, F_{m-1}(x_i))}{\partial F_{m-1}(x_i)}\right). \tag{8}$$

The final prediction $\hat{y}_i$, is the label that has the highest probability, i.e.,

$$\hat{y}_i = arg \max_c P(y_i = c), \tag{9}$$

where $P(y_i = c_1) = \frac{e^{F_M(x_i, c_1)}}{e^{F_M(x_i, c_1)} + e^{F_M(x_i, c_2)}} \tag{10}$

$$P(y_i = c_2) = \frac{e^{F_M(x_i, c_2)}}{e^{F_M(x_i, c_1)} + e^{F_M(x_i, c_2)}} \tag{11}$$

In the above, $F_M(x_i, z) = P\{F_M(x_i) = z\}, z \in \{c_1, c_2\}$, can be calculated by Eq.(1)-Eq.(8).

## 3.3 Base Detectors in GTB-RD Algorithm

When regression trees are used as base detectors, gradient boosting is referred to as gradient tree boosting (GTB), e.g., XGBoost and LightGBM [13] are all variants of GTB. In this paper, we adopt the GTB approach to rumor detecting, based on which an algorithm, GTB-RD algorithm, is developed.

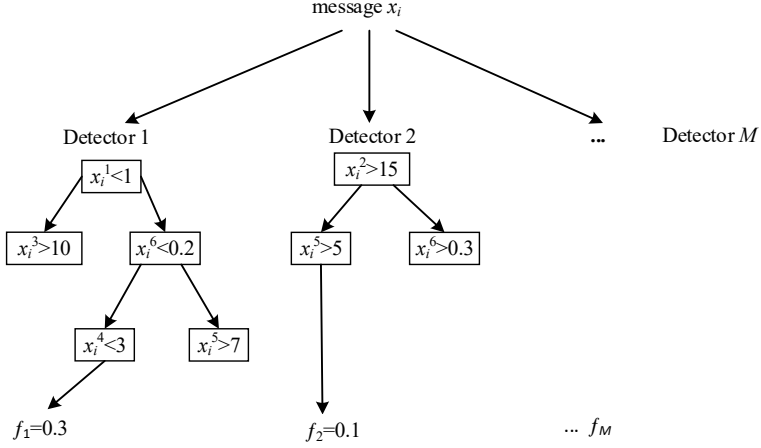

Fig. 1. Calculation of output prediction $F_M(x_i, z)$ when base detectors are regression trees.

In GTB-RD algorithm, $F_M(x_i, z)$ can be calculated as illustrated in Fig.1, which shows $M$ regression trees (detectors) that have been trained. Given an unlabeled message $x_i$, the predictions from base detectors are $f_1, f_2, \ldots, f_M$, respectively, and $F_M(x_i, z)$ is calculated as

$$F_M(x_i, z) = \gamma_1 f_1 + \gamma_2 f_2 + \cdots + \gamma_m f_M . \tag{12}$$

Then the predicted label $\hat{y}_i$ can be calculated by Eq.(9).

The training of a regression tree is as follows (the detailed algorithm is given as Algorithm 2 in Section 3.5). A regression tree $t(x)$ recursively splits the training samples $(x_1, y_1), (x_2, y_2), (x_3, y_3), \ldots, (x_n, y_n)$ into $M$ regions, $R_1, R_2, \ldots, R_M$, i.e.,

$$t(x) = \sum_{m=1}^{M} c_m I(x \in R_m), \tag{13}$$

$$\text{where } I(x \in R_m) = \begin{cases} 1, & \text{if } x \in R_m \\ 0, & \text{otherwise} \end{cases} \tag{14}$$

For each region of the tree, the output target is the average of the samples that the region contains, i.e.,

$$\hat{c}_m = \frac{1}{|R_m|} \sum_{x_i \in R_m} y_i \tag{15}$$

The training process of a regression tree is to find a splitting variable $j$ from the input features and a splitting point $s$ such that the samples fall into two regions $R_{left}$ and $R_{right}$:

$$R_{left}(j, s) = \{x | x^j \leq s\} \tag{16}$$

$$R_{right}(j, s) = \{x | x^j > s\} \tag{17}$$

For a message that falls into the region $R_{left}$ or $R_{right}$, the prediction is the averaged targets of the samples that the region contains:

$$\hat{c}_{left} = \frac{1}{|R_{left}(j,s)|} \sum_{x_i \in R_{left}(j,s)} y_i \tag{18}$$

$$\hat{c}_{right} = \frac{1}{|R_{right}(j,s)|} \sum_{x_i \in R_{right}(j,s)} y_i \tag{19}$$

Therefore, the squared loss of the samples is:

$$\sum_{x_i \in R_{left}(j,s)} (y_i - \hat{c}_{left})^2 + \sum_{x_i \in R_{right}(j,s)} (y_i - \hat{c}_{right})^2 \tag{20}$$

Finally, the optimization objective for each split is

$$\min_{j,s} \left[ \min_{c_{left}} \sum_{x_i \in R_{left}(j,s)} (y_i - \hat{c}_{left})^2 + \min_{c_{right}} \sum_{x_i \in R_{right}(j,s)} (y_i - \hat{c}_{right})^2 \right] \tag{21}$$

A regression tree stops splitting until it reaches the depth $P$ or the number of training messages that fall into the region is smaller than $H$. $P$ and $H$ are hyper-parameters that are defined in advance to prevent the regression tree from over-fitting.

### 3.4 Detecting deadline

The time when the rumor detecting is carried out is crucial. In general, the earlier to detect a message, the lower detecting accuracy would be. The time elapsed from the time when a message is posted to the time when the detecting is conducted is referred to as detecting deadline [3][5][10], denoted by $T$ in this paper.

For a sample $x_i$ in the given dataset $\mathcal{D}$, rearrange $x_i$ in the following format:

$$x_i = (const\ features; changing\ features) = (x_i^1, x_i^2, \ldots, x_i^t; x_i^{t+1}(T), x_i^{t+2}(T), \ldots, x_i^{t+p}(T)), \quad (22)$$

where $x_i^j$ ($j$=1,2,…,$t$) is the $j$:th constant feature; $x_i^k(T)$ ($k$=$t$+1,$t$+2,…,$t$+$p$) is the $k$:th changing feature. A constant feature is a feature whose value stays unchanged during the message propagation, e.g., "Message length"; while a changing feature is a feature whose value changes during the message propagation, e.g., "Num of comments". Define

$$V(x_i^k, T) = x_i^k(T), \quad (23)$$

where $V(x_i^k, T)$ represents the value of the $k$:th changing feature for message $x_i$ when detecting deadline is $T$.

### 3.5 The GTB-RD algorithm

Finally, the proposed GTB-RD algorithm is presented in Table 1.

Table 1. The GTB-RD Algorithm

| **Algorithm 1:** GTB-RD Algorithm |
|---|
| Given a dataset $D$ with $n$ samples, defined by Eq. (22). |
| Input: $T$: detecting deadline, $M$: number of base detectors, $\alpha$: learning rate, $v$: an unlabeled message to predict. |
| 1: **For** each sample in $D$ **do** |
| 2:    **For** each changing feature in sample $x_i$, **do** |
| 3:       Calculate feature value at detecting deadline $T$ by Eq. (23). |
| 4: Initialize $F_0$ to a constant. |
| 5: **For** $m$=1 to $M$ **do** |
| 6:    **For** each sample **do** |
| 7:       Update the target of each sample by Eq. (7). |
| 8:    **End** |
| 9:    Fit a regression tree to predict the targets by **Algorithm 2.** |
| 10:    Compute weight of the regression tree by Eq. (8). |
| 11:    Update the model by Eq. (3). |
| 12: **End** |
| 13: Obtain $F_M(x_i, c_1)$ by Eq. (2). |
| 14: Repeat step 4 to step 13 to obtain $F_M(x_i, c_2)$. |
| 15: Output: the predicted label for message $v$ by Eq. (9). |

Table 2 presents the training algorithm of base detectors in GTB-RD algorithm.

Table 2. Training algorithm of a base detector in GTB-RD algorithm

| **Algorithm 2:** Regression Tree Training |
|---|
| Input: $P$: maximum tree depth |
| 1:   **For** every feature in $x$ **do** |
| 2:       **For** every split point in current feature **do** |
| 3:           Calculate loss by Eq. (20). |
| 4:       **End** |
| 5:   **End** |
| 6:   Find the feature and the split point that minimize current loss. |
| 7:   Split the data into two regions by Eq. (16) and Eq. (17). Calculate the predication by Eq. (18) and Eq. (19). |
| 8:   Repeat step 1 to step 7, until the maximum depth $P$ is reached. |
| 9:   Output: the regression tree as Eq. (13). |

## 4   FEATURE CONSTRUCTION AND SELECTION

To any rumor detecting algorithm, features used in the algorithm are quite important and could influence the results obtained significantly [3][9]. In this Section, feature construction and selection are conducted.

    Feature construction: In related research, normally 20-30 features are used for Weibo rumor detecting [3][5][9]. In this paper, we adopt features used in related research and also propose new features. All these features are candidates for later feature selection. Table 3 presents 34 features that are used in later selection. Among them, 21 features are those used in the literature and 13 features are proposed in this paper.

Table 3. List of the features constructed for later selection

| Features used in related research | Features proposed in this paper |
|---|---|
| Time span, Num of '@'s, Num of topics, Num of emoji, Num of friends, Num of followers, Registration time, User influence, Num of comments, Num of reposts, User id, User name, User description, City, Province, User verified, Number of images, Number of videos, User gender, User location, Hyperlink | Message length, Num of question mark, Num of exclamation, Num of quotes, Num of brackets, First person pronouns, Second person pronouns, Third person pronouns, Num of dates, Num of digits, Num of bi-followers, Num of all messages, Num of likes |

    Feature selection: A feature selection algorithm is developed, as shown in Table 4. The most important features are identified and other features are dropped. It can be noted that due to the use of the GTB-based approach, feature importance can be relatively easily calculated, i.e., by how many times the feature is used for splitting in the tree growth. The more often a feature is used as the split point of a tree, the more important the feature is. For $n$ features $(x^1, x^2, ..., x^n)$, if there are $k$ regression trees $(h_1, h_2, ..., h_k)$ finally obtained, then the importance of feature $x^i$ is calculated by

$$feature\ importance\ (x^i) = \frac{\sum_{j=1}^{k} c_j^i}{\sum_{j=1}^{k} \sum_{i=1}^{n} c_j^i}, \quad (24)$$

where $c_j^i$ is the times that feature $x^i$ is used to split regression tree $h_j$.

Table 4. Feature selection algorithm

| | Algorithm 3: Feature Selection |
|---|---|
| | Input: $M$: number of base detectors, $\alpha$: learning rate, $P$: maximum tree depth, $u$: number of candidate features, $u_1$: number of features to be selected |
| 1: | Initialize $F_0$ to be a constant. |
| 2: | **For** $m=1$ to $M$ **do** |
| 3: |    Update predicting target by Eq.(7). |
| 4: |    Train a base detector by **Algorithm 2**. |
| 5: | **End** |
| 6: | **For** $i=1$ to $u$ **do** |
| 7: |    **For** $m=1$ to $M$ **do** |
| 8: |      Calculate the times that feature $x^i$ is used in base detector $m$. |
| 9: |    **End** |
| 10: |    Calculate feature importance by Eq.(24). |
| 11: | **End** |
| 12: | Sort the $u$ candidate features in a descending order of feature importance. |
| 13: | Output: Select the first $u_1$ features. |

Using the feature selection algorithm, 23 features are selected for Weibo rumor detecting, which are summarized in Table 5. The 23 features can be divided into three categories: message features, user features and comment features.

Table 5. Twenty-three features identified by feature selection (those with asterisks are features proposed in this paper)

| Category | Feature | Feature description |
|---|---|---|
| Message | Time span | The time interval between user registration and posting |
| | Message length* | Number of words and characters the message contains |
| | Num of question mark* | Number of symbol '?' the message contains |
| | Num of exclamation* | Number of symbol '!' the message contains |
| | Num of quotes* | Number of quotes symbol the message contains |
| | Num of brackets* | Number of brackets, including '()','[]','【】' and '{}' |
| | First person pronouns* | The number of first person pronouns |
| | Second person pronouns* | The number of second person pronouns |
| | Third person pronouns* | The number of third person pronouns |
| | Num of '@'s | '@' represents one user is mentioned in the message |
| | Num of topics | On Weibo, a topic is surrounded by the symbol '#' |
| | Num of dates* | The number of dates the message contains |
| | Num of digits* | The number of digits the message contains |
| | Num of emoji | The number of emoji the message contains |
| User | Num of friends | Number of people the author is following at time of post |
| | Num of followers | Number of people following the author at time of post |
| | Num of bi-followers* | Number of followers the author follows |
| | Registration time | The time of author registration |
| | Num of all messages* | Number of messages the user posted since registration |
| | User influence | Calculated by Eq. (25). |
| Comments | Num of comments | Number of comments on the original message |

| | Num of reposts | Number people who forward the message |
|---|---|---|
| | Num of likes[*] | How many times the message is liked |

$$User\ influence = \log\left(\frac{followers\_count - bi\_followers\_count}{friends\_count + 1}\right) \quad (25)$$

## 5 EXPERIMENTS AND RESULTS

In this Section, several experiments are conducted on a Weibo dataset. This is the only published dataset for rumor detecting on Weibo, and related research [3][5][6][7][8][10] all used this dataset to validate their Weibo rumor detecting algorithms proposed[4]. For comparative purpose, we also use this dataset in this paper to compare the proposed GTB-RD algorithm with existing algorithms.

### 5.1 Dataset used in the experiments

The dataset is provided by Sina community management center [15], which is an official debunking platform of Sina that reports known rumors. The dataset contains 2,313 labeled rumors and 2,351 labeled non-rumors. Table 6 provides the details of the dataset.

Table 6. Statistics of the dataset

| Statistic | Details |
|---|---|
| Num of users | 2,746,818 |
| Num of total comments | 3,805,656 |
| Num of events | 4,664 |
| Num of rumors | 2,313 |
| Num of non-rumors | 2,351 |
| Avg. time length / event | 2,460.7 Hours |
| Avg. num of comments / event | 816 |
| Max num of comments / event | 59,318 |
| Min num of comments / event | 10 |

### 5.2 Experiments and results

We compare the proposed detecting algorithm, GTB-RD algorithm, with the following state-of-the-art detecting algorithms:

  RNN-based algorithms [10]: Deep learning model RNN is used to capture the dynamic temporal characteristics of rumor diffusion given the sequential nature of text streams in social media, and four RNN-based rumor detecting algorithms, i.e., LSTM-1, GRU-1, GRU-2, and Tanh-RNN were proposed in [10].

  DT-Rank [3]: Decision-tree-based ranking algorithm used to identify rumors, which searches for inquiry phrases and clusters disputed factual claims, and ranks the clustered results based on statistical features.

  SVM-TS [5]: Linear SVM detecting algorithm that uses time-series structures to model the variation of social contents, users and propagation patterns.

---
[4] Some of the related research also conducted experiments on Twitter datasets (but those datasets are not published); but for rumor detecting on Weibo, all related research only used this dataset.

RFC [8]: Random forest classifier that uses three parameters to fit the temporal volume curve.

DTC [6] and SVM-RBF [7]: The algorithms use decision tree classifier and the SVM-based detecting algorithm with RBF kernel, both of which use hand-crafted features based on the overall statistics of the posts.

The proposed GTB-RD algorithm: The algorithm is implemented using python 3 and the selected 23 features are used in the algorithm.

The performance criteria are *accuracy*, *precision*, *recall* and $F_1$, which have also been used in related research [3][5][6][7][8][10] to evaluate the performance of detecting algorithms. These criteria are defined as follows.

$$accuracy = \frac{TP+TN}{TP+FP+TN+FN} \qquad (26)$$

$$precision = \frac{TP}{TP+FP} \qquad (27)$$

$$recall = \frac{TP}{TP+FN} \qquad (28)$$

$$F_1 = \frac{2 \times precision \times recall}{precision + recall} \qquad (29)$$

In the above, TP is True Positive, FP is False Positive, TN is True Negative and FN is False Negative.

All experiments are conducted on a PC with 3.30GHz CPU and 8G RAM. The proposed GTB-RD algorithm is run by 10-fold cross-validation for 10 times and the average values of accuracy, precision, recall and $F_1$ are presented in the last row of Table 7. It can be seen that GTB-RD has the best performance in all four criteria. As the related research which we compare with, [3][5][6][7][8][10], does not report the computation times used, it would be difficult to carry out a comparative study on computation times. However, the computation time of the GTB-RD algorithm is 91 seconds (the average value of 10 runs), which seems to be acceptable in practice.

Table 7. Rumor detecting results (the number of base detectors is 500, the maximum tree depth is 6, the learning rate is 0.2)

| Detecting algorithm | Class | Accuracy | Precision | Recall | $F_1$ |
| --- | --- | --- | --- | --- | --- |
| DT-Rank | R N | 0.732 | 0.732 | 0.732 | 0.731 |
| DTC | R N | 0.731 | 0.732 | 0.730 | 0.730 |
| RFC | R N | 0.772 | 0.794 | 0.771 | 0.768 |
| SVM-RBF | R N | 0.818 | 0.819 | 0.818 | 0.818 |
| SVM-TS | R N | 0.857 | 0.859 | 0.858 | 0.859 |
| Tanh-RNN | R N | 0.873 | 0.886 | 0.873 | 0.873 |
| LSTM-1 | R N | 0.896 | 0.900 | 0.913 | 0.908 |
| GRU-1 | R N | 0.908 | 0.912 | 0.908 | 0.908 |
| GRU-2 | R N | 0.910 | 0.914 | 0.910 | 0.910 |
| GTB-RD | R N | **0.951** | **0.934** | **0.970** | **0.951** |

**5.3 Study on the influence of hyper-parameters**

It is known that hyper-parameters in a model or an algorithm could have significant influence on the performance of the model or algorithm; therefore, it would be interesting to study how much influence the hyper-parameters may have on the performance of the proposed GTB-RD algorithm.

In GTB-RD algorithm, there are three important hyper-parameters, i.e., the number of base detectors ($M$ in Eq.(2)), the maximum tree depth ($P$ in Algorithm 2), the learning rate ($\alpha$ in Eq.(3)). We use a way similar to that of sensitivity analysis in [17] to study the influence of hyper-parameters on the accuracy. The influences of hyper-parameters on precision, recall and $F_1$ can be studied in a similar manner, which are omitted due to the length limit of the paper.

Number of base detectors (*M*): Fig.2 shows how the detecting accuracy of GTB-RD algorithm changes with different *M* values. We study the cases for different maximum tree depth, *P*=5, 10, 20, respectively. It can be seen that in all three cases, the influence of *M* values on the accuracy is quite small. For example, for GTB-RD algorithm of *P*=5, when *M* changes from 20 to 200, the accuracy only fluctuates between 91.96% and 93.78%, i.e., a maximum of 1.82% absolute change or 1.98% relative change.

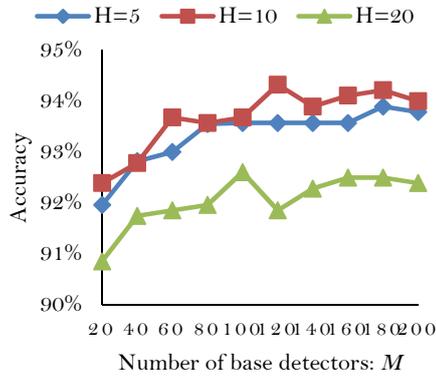 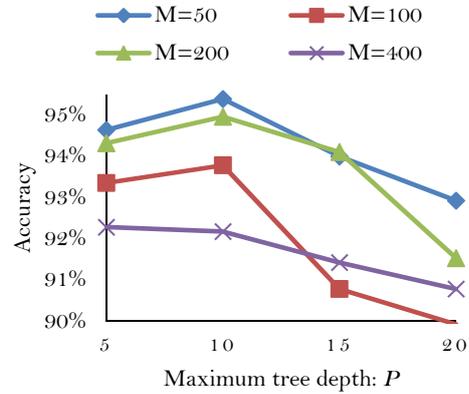

Fig.2. Influence of number of base detectors.  Fig.3. Influence of maximum tree depth.

Maximum tree depth (*P*): Fig.3 shows how the detecting accuracy of GTB-RD algorithm changes with different *P* values. We study the cases for different number of base detectors, *M*=50, 100, 200, 400, respectively. A similar small influence of *P* values is observed. For example, for GTB-RD algorithm of *M*=400, when *P* changes from 5 to 20, the accuracy only fluctuates between 92.28% and 90.78%, i.e., a maximum of 1.50% absolute change or 1.63% relative change.

Learning rate ($\alpha$): Fig.4 shows how the detecting accuracy of GTB-RD algorithm changes with different $\alpha$ values. We study the cases for different *M* and *P* values. A similar small influence of $\alpha$ values is observed. For example, for GTB-RD algorithm of *P*=5 and *M*=100, when $\alpha$ changes from 0.1 to 1, the accuracy only fluctuates between 95.50% and 94.32%, i.e., a maximum of 1.18% absolute change or 1.24% relative change.

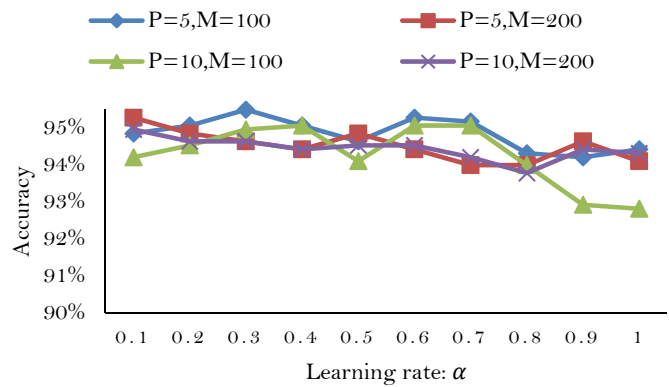

Fig.4. Influence of learning rate.

It seems from the above results that in general the influence of the hyper-parameters on the accuracy of the proposed GTB-RD algorithm is relatively small, which suggests that the GTB-RD algorithm is relatively robust.

In Table 7, the results of the GTB-RD algorithm (i.e., the last row) are obtained under the best values of hyper-parameters; thus it would be interesting to investigate what if the values of hyper-parameters are not chosen to be the best ones. To illustrate, we choose the number of base detectors, maximum tree depth, and learning rate to be the third-best values, and the results obtained are shown in Table 8. It can be seen that under this set of not-so-good hyper-parameters, the proposed GTB-RD algorithm still has the best performance, which partially validates the usefulness of the proposed GTB-RD algorithm.

Table 8. Rumor detecting results (number of base detectors is 100, maximum tree depth is 20, learning rate is 0.3)

| Detecting algorithm | Class | Accuracy | Precision | Recall | $F_1$ |
|---|---|---|---|---|---|
| DT-Rank | R N | 0.732 | 0.732 | 0.732 | 0.731 |
| DTC | R N | 0.731 | 0.732 | 0.730 | 0.730 |
| RFC | R N | 0.772 | 0.794 | 0.771 | 0.768 |
| SVM-RBF | R N | 0.818 | 0.819 | 0.818 | 0.818 |
| SVM-TS | R N | 0.857 | 0.859 | 0.858 | 0.859 |
| Tanh-RNN | R N | 0.873 | 0.886 | 0.873 | 0.873 |
| LSTM-1 | R N | 0.896 | 0.900 | 0.913 | 0.908 |
| GRU-1 | R N | 0.908 | 0.912 | 0.908 | 0.908 |
| GRU-2 | R N | 0.910 | 0.914 | 0.910 | 0.910 |
| GTB-RD | R N | **0.923** | **0.919** | **0.927** | **0.923** |

## 5.4 Early rumor detecting

In this Section, we study the problem of early rumor detecting, i.e., the detecting accuracy an algorithm can achieve when detecting deadline $T$ is small. Fig.5 and Table 9 depict the detecting accuracy under different $T$ values. From Fig.5 it can be seen that under any value of $T$, the proposed GTB-RD algorithm has the highest detecting accuracy; specifically, the advantage of the GTB-RD algorithm for early rumor detecting, i.e., when $T$ is small, is quite clear. Table 9 presents the accuracy of different detecting algorithms when $T$ is small. For example, even when $T$=0, the proposed GTB-RD algorithm still has a relatively high detecting accuracy of 89.1%.

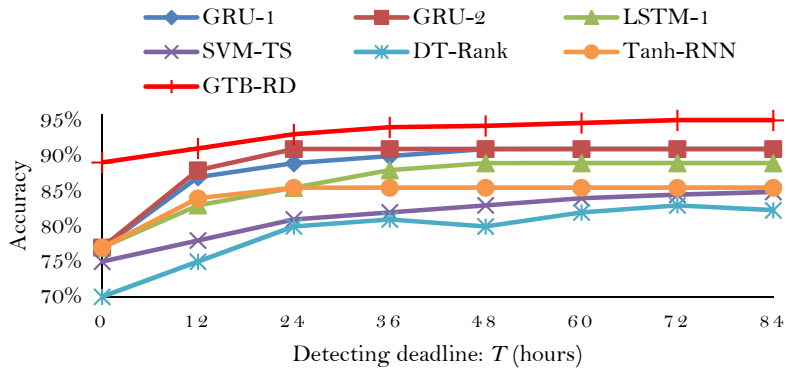

Fig.5. Rumor detecting accuracy under different *T*.

Table 9. Rumor detecting accuracy under small *T* (Hours)

| Detecting algorithm | T=0 | T=4 | T=8 | T=12 | T=24 |
|---|---|---|---|---|---|
| DT-Rank | 0.698 | 0.715 | 0.736 | 0.750 | 0.801 |
| SVM-TS | 0.750 | 0.781 | 0.764 | 0.782 | 0.812 |
| Tanh-RNN | 0.772 | 0.783 | 0.831 | 0.841 | 0.855 |
| LSTM-1 | 0.761 | 0.782 | 0.812 | 0.834 | 0.851 |
| GRU-1 | 0.770 | 0.825 | 0.847 | 0.873 | 0.893 |
| GRU-2 | 0.771 | 0.836 | 0.867 | 0.881 | 0.912 |
| GTB-RD | **0.891** | **0.907** | **0.917** | **0.911** | **0.931** |

## 6 CONCLUSIONS

This paper studies the problem of rumors detecting on the Chinese microblogging platform, Sina Weibo. In this paper, we proposed to use GTB approach to rumor detecting, based on which the GTB-RD algorithm is developed. Besides high detecting accuracy, another advantage that GTB-based approach brings is that feature selection can be done in a convenient way. In this paper, a feature selection algorithm is developed. Experiments on a widely used Weibo dataset show that the proposed GTB-RD algorithm outperforms the state-of-the-art detecting algorithms; moreover, it has the highest detecting accuracy at the early stage.

In this paper, we only conducted experiments on a Weibo dataset due to the availability of datasets. Nevertheless, GTB-based approach could be applicable to rumor detecting on other microblogging platforms as well, e.g., Twitter, and could result in better performance as is observed for Weibo in this paper. Feature selection can also be conducted using the algorithm proposed in this paper. In our future work, we will try to obtain datasets on microblogging platforms other than Weibo to validate more thoroughly the advantages of adopting GTB-based approach in rumor detecting.